\documentclass[10pt, aps, pra, a4paper, superscriptaddress, notitlepage, reprint, amsfonts, amsmath, amssymb, longbibliography]{revtex4-1}
\pdfoutput=1
\usepackage[utf8]{inputenc}
\usepackage{graphicx}
\usepackage{epstopdf}
\usepackage{xcolor}
\usepackage{physics}
\usepackage{hyperref}

\newcommand{\ii}{\mathrm{i}}
\newcommand{\ee}{\mathrm{e}}

\newcommand{\rrt}{\left(\vb*{r},t\right)}
\newcommand{\rr}{\left(\vb*{r}\right)}

\begin{document}
\title{Dispersion relations and self-localization of quasiparticles in coupled elongated Bose-Einstein condensates}
\author{M. R. Momme}
\affiliation{Physikalisch-Technische Bundesanstalt (PTB), Bundesallee 100, D-38116 Braunschweig, Germany}
\affiliation{Technische Universität Braunschweig, D-38106 Braunschweig, Germany}
\author{O. O. Prikhodko}
\affiliation{Department of Physics, Taras Shevchenko National University of Kyiv, Volodymyrska Str. 64/13, Kyiv 01601, Ukraine}
\author{Y. M. Bidasyuk}
\affiliation{Physikalisch-Technische Bundesanstalt (PTB), Bundesallee 100, D-38116 Braunschweig, Germany}

\date{\today}

\begin{abstract}
We present a detailed study of the spectrum and dispersion of Bogoliubov quasiparticles in two coupled elongated Bose-Einstein condensates.
We develop an analytically solvable model that approximates two infinite homogeneous condensates and compare its predictions to a numerical simulation of a realistic trapped system. 
While the comparisons show a reasonable agreement between the two models, they also manifest the existence of several anomalous Bogoliubov modes in the spectrum. 
These modes show degeneracy in both energy and momentum together with self-localization in the coordinate space.
\end{abstract}

\maketitle

\section{Introduction}
Systems consisting of mutually coherent Bose-Einstein condensates (BECs) have attracted considerable research interest in recent years.
While the most prominent applications of such systems are undoubtedly  matter-wave interferometry and quantum metrology \cite{RevModPhys.81.1051,RevModPhys.90.035005}, they also 
provide an excellent platform to study various other physical phenomena, including Josephson effects \cite{PhysRevA.59.620,PhysRevLett.95.010402,PhysRevLett.106.025302, PhysRevA.98.043624,PhysRevLett.105.204101}, quantum fluctuations and spatial coherence \cite{PhysRevLett.87.050404, PhysRevLett.87.160406,Hofferberth2007}, 
 and spin-orbital coupling \cite{Lin2011,Zhai2015}.
To enable such experiments, a coherent coupling between the condensates can be achieved in different ways.
Most common realizations utilize two internal states of atoms in BEC coupled by Raman lasers \cite{Abad2013,PhysRevLett.105.204101}, or multi-well trapping potentials with a possibility for atoms to tunnel through the barriers \cite{PhysRevLett.95.010402,PhysRevLett.106.025302}.
However, the physics of coupled condensates is determined not only by the nature of the coupling but also to a large extent by the geometry of individual condensates (see e.g. \cite{PhysRevLett.124.045301,PhysRevA.94.033603,Oliinyk_2019,Luick2020}).
One of the most simple yet nontrivial geometries is realized with two parallel cigar-shaped condensates coupled through a potential barrier along their long dimension. 
Despite their seeming simplicity, such systems display a wide range of dynamical effects with many open questions.
Having many similarities with superconducting long Josephson junctions \cite{PhysRevA.100.033601,PhysRevA.71.011601} such systems are actively studied in the context of sine-Gordon solitons and Josephson vortices \cite{PhysRevA.71.011601,PhysRevA.93.033618,10.21468/SciPostPhys.4.3.018}.
Coupled condensates with such geometry are in the focus of the present work.

Near-equilibrium dynamics of a Bose-condensed system is commonly analyzed on the level of low-energy collective excitations, also termed Bogoliubov quasiparticles. 
The spectrum of such collective excitations is very sensitive to the geometry of the system and reveals a number of specific features for the case of two coupled condensates \cite{PhysRevA.55.2935,PhysRevA.64.013615,PhysRevA.68.053609,PhysRevA.67.023606,Abad2013}.
In particular, the tunneling of atoms through the barrier may be effectively coupled with their motion inside each condensate if corresponding collective modes posses similar energies. 
The signatures of such a coupling were previously identified in theoretical \cite{PhysRevA.100.033601,PhysRevA.94.033603,Bouchoule2005} and experimental studies \cite{PhysRevLett.106.025302}. 
However, an understanding of general requirements for such coupled modes to appear, as well as their structure and dynamical properties, is still lacking.
In the present work we aim to develop such an understanding by extracting and analyzing the spectrum of linear collective excitations and building the dispersion relations of quasiparticle modes in parallel coupled elongated BECs.
To this end we develop an analytical model, which approximates two coupled infinite homogeneous condensates and which can be compared to the numerical results without fitting parameters.
A comparison of the analytical predictions to the numerical calculations of a realistic trapped  system is the main goal of the present work. 
Such a comparison allows to identify and analyze a peculiar phenomenon of self-localization, which is observed for some of the quasiparticle modes as a drastic deviation from the analytically predicted behavior.
As we show here, these modes correspond to a coupled internal and mutual motion in two condensates 
and can be associated with imaginary-wavenumber solutions of the analytical model.

The article is structured as follows. In section \ref{sec:Analytics} we derive and analyze the dispersion relation for linear collective excitations in a system of two parallel coupled homogeneous condensates. 
The more realistic setting of a finite trapped condensate is introduced in section \ref{sec:Trapped}.
We calculate the frequency and momentum spectrum of the Bogoliubov quasiparticles numerically and compare the results to the analytical predictions. 
Finally, in section \ref{sec:Modes} we identify and discuss several anomalous low-lying modes which appear outside of the predicted dispersion branches and display spatial self-localization. 

\section{Collective excitations in homogeneous infinite condensates}\label{sec:Analytics}
We consider an atomic Bose-Einstein condensate characterized by the mean-field Gross-Pitaevskii equation~(GPE)
\begin{equation}\label{eqn:GPE}
	\ii\hbar\pdv{t}\Psi\rrt = \left( -\frac{\hbar^2 \laplacian}{2M} + V\rr + g \abs{\Psi}^2\right) \Psi\rrt,
\end{equation}
with the nonlinear coupling coefficient $g>0$ to ensure stability of the condensate.
Our system of interest consists of two elongated weakly coupled BECs. 
A schematic representation of the system is depicted in Fig.~\ref{fig:sketch}.
In order to make an analytical treatment possible we consider in this section a uniform condensate in $x$-direction and a two-well trap in the transverse $(y,z)$-plane. We therefore assume $V = V_{\perp}(y,z)$. The specific shape of the two-well potential $V_{\perp}$ is not relevant here. 
It is only important that the total condensate wave function can be approximately represented as a sum of two parts, one for each condensate, and the dimensions can be separated into the dynamical ($x$) and the frozen ones ($y$,$z$):
\begin{figure}[tbp]
	\includegraphics[width=0.9\linewidth]{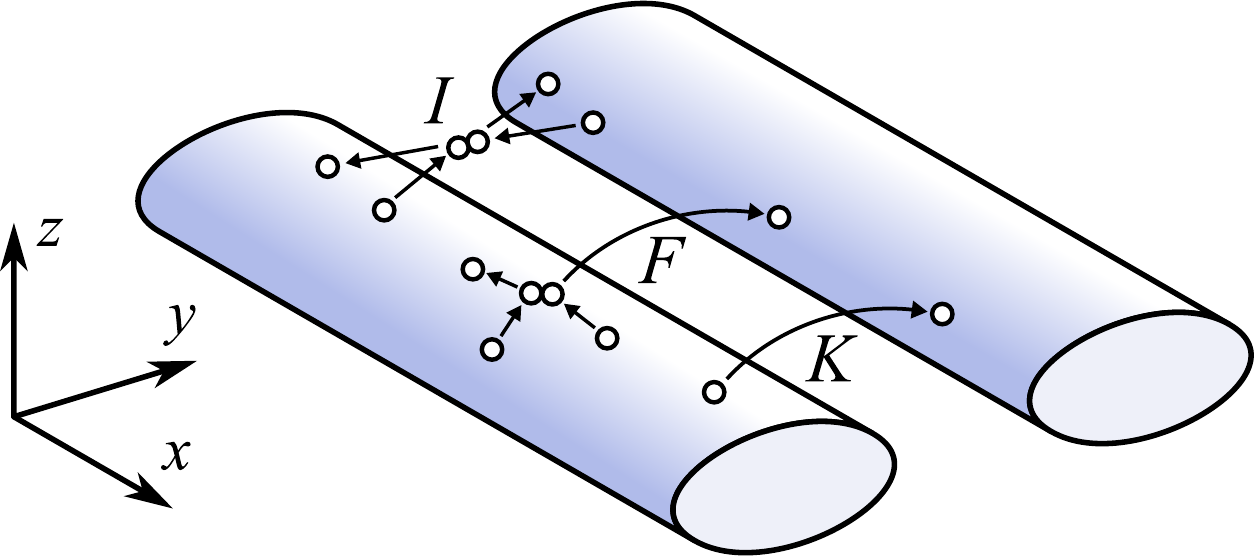}
	\caption{Schematic representation of coupled condensates in our geometry. Two elongated condensates are uniform in the $x$-direction and coupled through a barrier in the $y$-direction. 
	A few examples of particle-exchange processes are depicted to illustrate the physical meaning of coupling terms $K$, $F$ and $I$ in Eq.~(\ref{eqn:1DC-GPE}). More details on these coupling coefficients can be found in the text.
	}
	\label{fig:sketch}
\end{figure}
\begin{equation}\label{eqn:1D-TMM-ansatz}
	\Psi(x,y,z,t) = \Psi_1(x,t) \chi_1(y,z) + \Psi_2(x,t) \chi_2(y,z).
\end{equation}
Each of the functions $\chi_1$ and $\chi_2$ is localized in one well of the two-well potential and represents each one of the two coupled condensates. 
We assume these functions to be real, orthogonal and normalized to unity 
\[
\iint\dd{y} \dd{z} \chi_1\chi_2 = 0, \quad \iint\dd{y}\dd{z} \chi_1^2 = \iint\dd{y}\dd{z} \chi_2^2 = 1.
\]
Also a weak coupling between the two condensates implies that the absolute overlap of these functions is small
\begin{equation}\label{eq:overlap}
\iint\dd{y} \dd{z} |\chi_1\chi_2| \ll 1.
\end{equation}
The ansatz (\ref{eqn:1D-TMM-ansatz}) is an extension to a well-known two-mode approximation \cite{PhysRevA.59.620}. 
The key difference is that we keep one spatial dimension as a dynamical variable.
Inserting this ansatz into the GPE (\ref{eqn:GPE}), the frozen directions can be integrated out. 
This results in two coupled equations for the functions $\Psi_1$ and $\Psi_2$. 
The first equation reads
\begin{align}\label{eqn:1DC-GPE}
	&\ii \hbar \pdv{t} \Psi_1(x,t) = \left(-\frac{\hbar^2}{2M} \frac{\partial^2}{\partial x^2} +  g_\mathrm{1D} \abs{\Psi_1}^2 \right) \Psi_1 - K \Psi_2 \notag\\
	&- F \left( (\abs{\Psi_1}^2 + \abs{\Psi_2}^2) \Psi_2 + (\Psi_1^\ast \Psi_2 + \Psi_2^\ast \Psi_1) \Psi_1 \right) \notag\\
	&+ I \left(\abs{\Psi_2}^2 \Psi_1 + (\Psi_1^\ast \Psi_2 + \Psi_2^\ast \Psi_1) \Psi_2\right),
\end{align}
and the second equation is the same with indices 1 and 2 interchanged. 
The coefficients that enter these equations are defined as
\begin{align}
	g_\mathrm{1D} &= g \iint\dd{y} \dd{z} \chi_1^4 = g \iint\dd{y}\dd{z} \chi_2^4, \label{eq:g1d-coef}\\
	K &= - \iint\dd{y}\dd{z} \left(-\frac{\hbar^2}{2m} \chi_1 \laplacian_{y,z} \chi_2 + \chi_1 V_{\perp} \chi_2 \right), \label{eq:K-coef}\\
	F &= - g \iint\dd{y}\dd{z} \chi_1^3 \chi_2 = - g \iint\dd{y}\dd{z} \chi_1 \chi_2^3, \label{eq:F-coef}\\
	I &= g \iint\dd{y}\dd{z} \chi_1^2 \chi_2^2. \label{eq:M-coef}
\end{align}
For simplicity we only consider the case of a fully symmetric two-well potential.
More general equations for asymmetric wells can be derived in the same way. 

The ground state of two coupled parallel condensates is characterized by a uniform particle density $n$ in each condensate. 
So we can write the ground state solution of Eq.~(\ref{eqn:1DC-GPE}) as $\Psi_1 = \Psi_2 = \sqrt{n}\, \ee^{-\ii\mu t/\hbar}$ with the chemical potential $\mu = -K + (g_\mathrm{1D} -4F +3I)n$.
In order to analyze collective excitations in the system we introduce a small perturbation to the ground state in a usual form of plane waves \cite{Pitaevskii2003},
\begin{equation}\label{eqn:perturbed_gs}
	\begin{aligned}
	\Psi_{1}(x,t) &= \ee^{-\ii\mu t/\hbar} \left(\sqrt{n} + u_{1}\ee^{-\ii\omega t + \ii k x}  + v_{1}^{\ast}\ee^{\ii\omega t-\ii k x}  \right),\\
	\Psi_{2}(x,t) &= \ee^{-\ii\mu t/\hbar} \left(\sqrt{n} +u_{2} \ee^{-\ii\omega t + \ii k x}  +v_{2}^{\ast} \ee^{\ii\omega t-\ii k x}  \right).
	\end{aligned}
\end{equation}
Inserting this into Eq.~(\ref{eqn:1DC-GPE}) yields the Bogoliubov-de Gennes system of four coupled equations, which can be expressed in the following matrix form
\begin{equation}\label{eq:bdg_2c}
\left[\hat L_0 + K \hat L_K + (g_\mathrm{1D} \hat L_g + F \hat L_F  + I \hat L_I) n \right]\!
\begin{bmatrix}
u_1 \\
v_1 \\
u_2 \\
v_2
\end{bmatrix}=
\hbar\omega
\begin{bmatrix}
u_1 \\
v_1 \\
u_2 \\
v_2
\end{bmatrix}
\end{equation} 
with

\[
\hat L_0 = \left(\frac{\hbar^2 k^2}{2M}- \mu\right) \begin{bmatrix}
1 & 0 & 0 & 0 \\
 0 & -1 & 0 & 0 \\
0 & 0 & 1 &  0 \\
0 & 0 & 0 & -1
\end{bmatrix},
\]

\[
\hat L_g = \begin{bmatrix}
2 & 1 & 0 & 0 \\
-1 & -2 & 0 & 0 \\
0 & 0 & 2 & 1 \\
0 & 0 & -1 & -2 
\end{bmatrix},\,\,\,
\hat L_K = \begin{bmatrix}
0 & 0 & -1 & 0 \\
0 & 0 & 0 & 1 \\
-1 & 0 & 0 & 0 \\
0 & 1 & 0 & 0 
\end{bmatrix},
\]

\[
\hat L_F = \begin{bmatrix}
-4 & -2 & -4 & -2 \\
2 & 4 & 2 & 4 \\
-4 & -2 & -4 & -2 \\
2 & 4 & 2 & 4 
\end{bmatrix},
\,\,\,
\hat L_I = \begin{bmatrix}
2 & 1 & 4 & 2 \\
-1 & -2 & -2 & -4 \\
4 & 2 & 2 & 1 \\
-2 & -4 & -1 & -2 
\end{bmatrix}.
\]

The eigenvalues of Eq.~(\ref{eq:bdg_2c}) can be calculated analytically and provide the dispersion relations, which form two separate branches
\begin{equation}\label{eq:branch1}
\left[\hbar\omega_1(k)\right]^2 = \frac{\hbar^2 k^2}{2 M} \left( \frac{\hbar^2 k^2}{2 M} + 2 g_\mathrm{1D} n - 8 F n + 6 I n\right),
\end{equation}

\begin{align}\label{eq:branch2}
& \left[\hbar\omega_2(k)\right]^2 = \left( \frac{\hbar^2 k^2}{2 M} + 2 K + 4 F n - 4 I n \right) \notag\\ & \times \left( \frac{\hbar^2 k^2}{2 M} + 2 g_\mathrm{1D} n + 2 K + 4 F n - 6 I n \right).
\end{align}

The frequency $\omega_1$ corresponds to excitations which are symmetric (in-phase) in the two condensates while $\omega_2$ corresponds to antisymmetric (out-of-phase) ones. 
These results are consistent with previous studies of coherently coupled spinor BECs in Refs.~\cite{PhysRevA.55.2935, Abad2013,PhysRevA.64.013615} in the limit of purely linear coupling ($F=0$, $I=0$). 
In spinor BECs the  branch $\omega_1(k)$ represents a density wave, and $\omega_2(k)$ is a spin wave. 
We will therefore refer to the corresponding excitations in a two-well condensate as \emph{density-like} and \emph{spin-like}.

The density-like branch $\omega_1$ is gapless and in the long-wavelength limit ($k\rightarrow 0$) it describes sound modes propagating with a characteristic velocity  
\begin{equation}\label{eq:speedofsound1}
c = \eval{\dv{\omega_1}{k}}_{k\rightarrow 0} = \sqrt{\frac{(g_\mathrm{1D} - 4 F + 3 I) n}{M}}.
\end{equation}

The spin-like branch $\omega_2$ is gapped, with the gap size corresponding to the frequency of  Josephson plasma oscillations \cite{PhysRevA.100.033601}. From Eq. (\ref{eq:branch2}) we find this frequency as
\[
\hbar \omega_p = 2 \sqrt{ \left( K + 2 F n - 2 I n \right) \left( g_\mathrm{1D} n + K + 2 F n - 3 I n \right)}.
\]
This expression is similar to the result of so-called full two-mode model presented in \cite{PhysRevA.95.023627}.

In order to get further understanding of the obtained dispersion relations, it is necessary to analyze the coefficients defined in Eqs.~(\ref{eq:g1d-coef})--(\ref{eq:M-coef}).
The coefficient $g_\mathrm{1D}$ represents the one-dimensional reduction of the nonlinear interaction parameter $g$. 
It corresponds to a collisional interaction between atoms inside each condensate.
The other three coefficients $K$, $F$ and $I$ characterize the coupling between the two condensates. 
Their mathematical form and physical meaning are very similar to the corresponding coefficients of the two-mode model \cite{PhysRevA.95.023627,PhysRevA.90.043610}. 
The linear coupling coefficient $K$ represents the probability density of a single particle tunneling through the potential barrier. 
The other two coefficients $F$ and $I$ and  corresponding terms in  Eq.~(\ref{eqn:1DC-GPE}) represent collisional coupling between the two condensates. 
From the integrals (\ref{eq:K-coef}--\ref{eq:M-coef}) we see also that $K$ and $F$ are both of first order with respect to the overlap between $\chi_1$ and $\chi_2$, while $I$ is of second order. 
This means that if Eq.~(\ref{eq:overlap}) holds then also
\begin{equation}
g_\mathrm{1D} n \gg (|K|, |F n|) \gg I n.
\end{equation} 
For many physically relevant applications the terms in Eq.~(\ref{eqn:1DC-GPE}) proportional to $I$ will be negligible, except for very strong interaction regimes, where they can significantly alter the dynamical phase portrait \cite{Liu_2010}.
Here we do not address the specific effects of these terms and consider $I=0$ for the rest of the paper.

From Eqs.~(\ref{eq:g1d-coef})--(\ref{eq:M-coef}) one may also see that two coefficients, $g_\mathrm{1D}$ and $I$, are strictly positive and the other two, $K$ and $F$, are in general not sign definite. 
Then the requirement that the frequency $\omega_2$ of spin-like modes (\ref{eq:branch2}) is real-valued imposes an additional restriction: 
\begin{equation}\label{eq:k2fn}
    K+2Fn>0.
\end{equation}
Otherwise the ground state becomes formally unstable which would show a general inconsistency of the proposed one-dimensional model.
In this context it is worth noticing that equations with purely linear coupling are often used in the modeling of two-well condensates \cite{PhysRevA.71.011601,PhysRevA.68.053609,Bouchoule2005}. 
However the requirement (\ref{eq:k2fn}) then reads simply as $K>0$, which greatly limits the applicability of such models or requires to treat the coefficient $K$ as a fitting parameter (see e.g. discussions in Refs.~\cite{PhysRevA.73.013604,PhysRevA.81.025602}).
We illustrate this issue and analyze the behavior of the coupling coefficients for a more specific physical system in the next section (see Fig.~\ref{fig:KF} below).

\section{Quasiparticle spectrum and dispersion in a finite system}\label{sec:Trapped}

Let us now turn to the more realistic case of a trapped system.
We consider a trap with the shape of a symmetric double-well potential in $y$-direction and a box-like potential in $x$-direction:
\begin{multline}\label{eqn:pot}
V(x,y,z) =  A\left[1+e^{\frac{1}{\lambda}(L/2-|x|)}\right]^{-1} \\ + \frac{M}{2}(\omega_y^2 y^2 + \omega_z^2 z^2) + V_b \,\ee^{-2 y^2 / \lambda^2},
\end{multline}
where $\omega_y = 2\pi\cdot 50\,$Hz and $\omega_z = 2\pi\cdot 200\,$Hz are the trap frequencies in transverse directions, $L=180\,\mu$m and $A/h = 1500\,$Hz are the length and the depth of the box-like potential in the longitudinal direction, $\lambda = 2\,\mu$m is the characteristic width of the barrier and the box potential edge, which in a real experiment would be related to the resolution of the optical system.  
We consider the amplitude of the barrier potential $V_b$ as a tunable parameter controlling the coupling between the two condensates.
While this trap configuration does not reproduce any specific experimental setup, such traps are accessible in present-day experiments with painted potentials \cite{Tajik:19} or atom chips \cite{PhysRevLett.106.025302}.

The trapped system is modeled by the GPE (\ref{eqn:GPE}), however for computational simplicity we reduce it to two spatial dimensions, while the $z$-dimension can be safely considered as frozen in the low-energy region that we aim to analyze. 
A usual dimensional reduction procedure leads to the rescaling of the interaction parameter $g$, which in the two-dimensional approximation becomes \cite{Bao2003}
\begin{equation*} \label{eqn:interaction_g}
	g = a \sqrt{\frac{8\pi \hbar^3 \omega_z}{M}},
\end{equation*}
where $a$ is the s-wave scattering length of the atoms. 
We consider here $^{87}\mathrm{Rb}$ with $M = 86.91~\mathrm{u}$ and $a = 5.313~\mathrm{nm}$. 
The total particle number is $N=5.5\cdot 10^4$ and it defines the normalization of the wave function.

The stationary ground state of the system $\psi_g$ and the corresponding chemical potential $\mu$ is obtained by propagating the GPE (\ref{eqn:GPE}) in imaginary time \cite{Bao2003}. 
In order to calculate the spectrum of collective excitations corresponding to the stationary state $\psi_g$ we solve a standard Bogoliubov -- de Gennes system of equations in two spatial dimensions
\cite{Pitaevskii2003}
\begin{equation}\label{eqn:BdG}
    \begin{aligned}
    \hbar\omega u = (\hat{H}_0 + 2g \abs{\psi_g}^2 -\mu) u +  g |\psi_g|^2 v, \\	
    -\hbar\omega v = (\hat{H}_0 + 2g \abs{\psi_g}^2 -\mu) v +  g |\psi_g|^2 u,
    \end{aligned}
\end{equation}
where
\begin{equation}
\hat{H}_0 = -\frac{\hbar^2\laplacian}{2M} + V\rr.
\end{equation}
The functions $u\rr$ and $v\rr$ represent the spatial distribution of a Bogoliubov mode with a characteristic frequency $\omega$. 

\begin{figure}[tbp]
	\includegraphics[width=\linewidth]{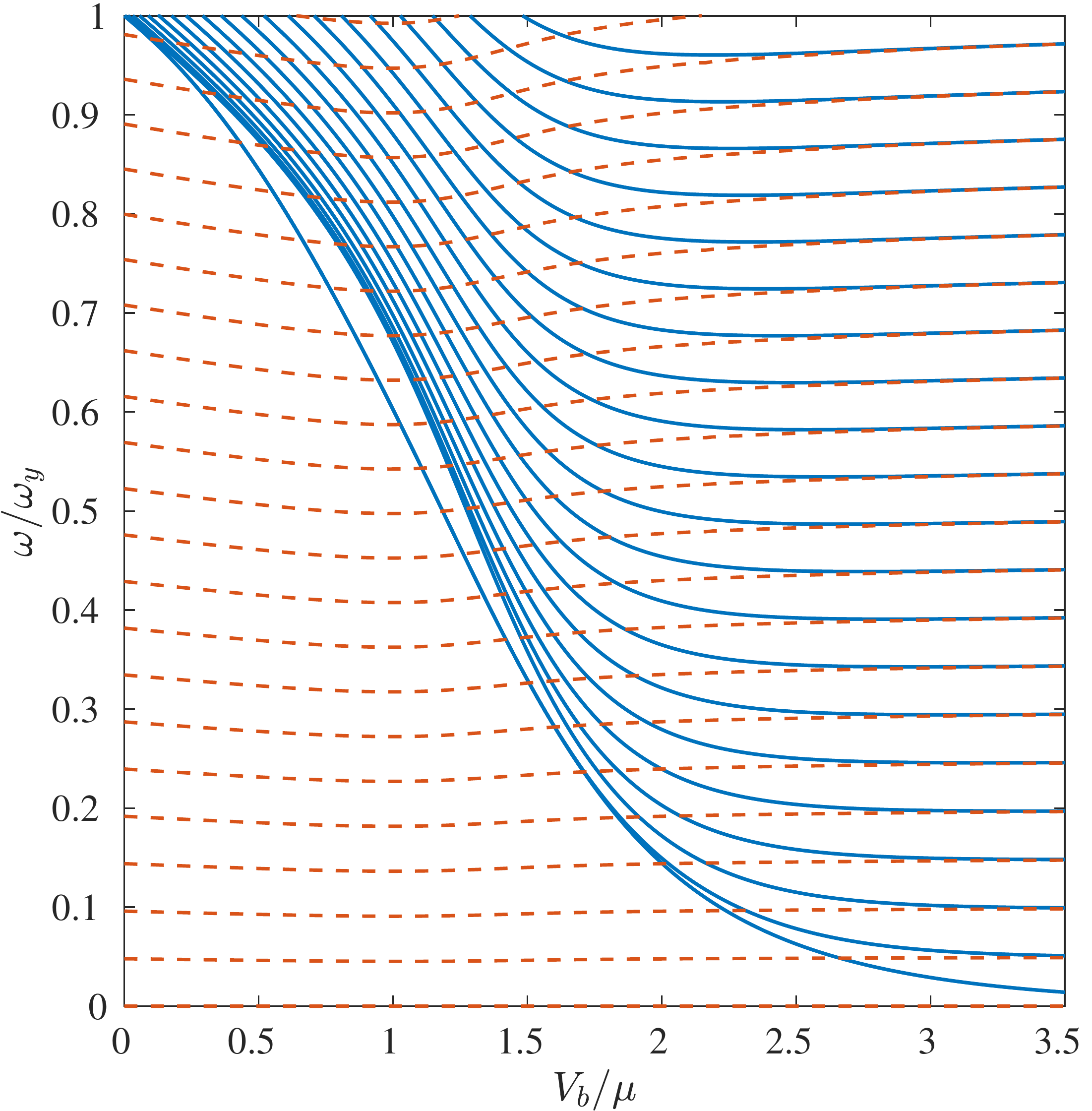}
	\caption{Numerically obtained spectrum of Bogoliubov excitations as a function of  $V_b/\mu$. The modes are distinguished by their symmetry: the red dotted lines correspond to density-like modes, blue solid lines correspond to spin-like modes.}
	\label{fig:BarrierScan}
\end{figure}

The numerically obtained spectrum of eigenfrequencies of Eqs.~(\ref{eqn:BdG}) is shown in Fig.~\ref{fig:BarrierScan} as a function of the dimensionless ratio $V_b/\mu$. 
We can distinguish two types of modes by their spatial symmetry across the barrier. 
Modes which are symmetric across the barrier can be identified as density-like excitations. 
We observe that they are rather insensitive to the barrier height. 
The other type of modes are antisymmetric across the barrier and they are identified as spin-like excitations. 
They are sensitive to the barrier height. 
In the limit of a vanishing barrier the lowest spin-like mode is a dipole mode with the characteristic frequency $\omega = \omega_y$. 
For very high barriers both types of modes become degenerate as they approach the spectrum of two uncoupled condensates. 

\begin{figure}[tbp]
    \centering
	\includegraphics[width=\linewidth]{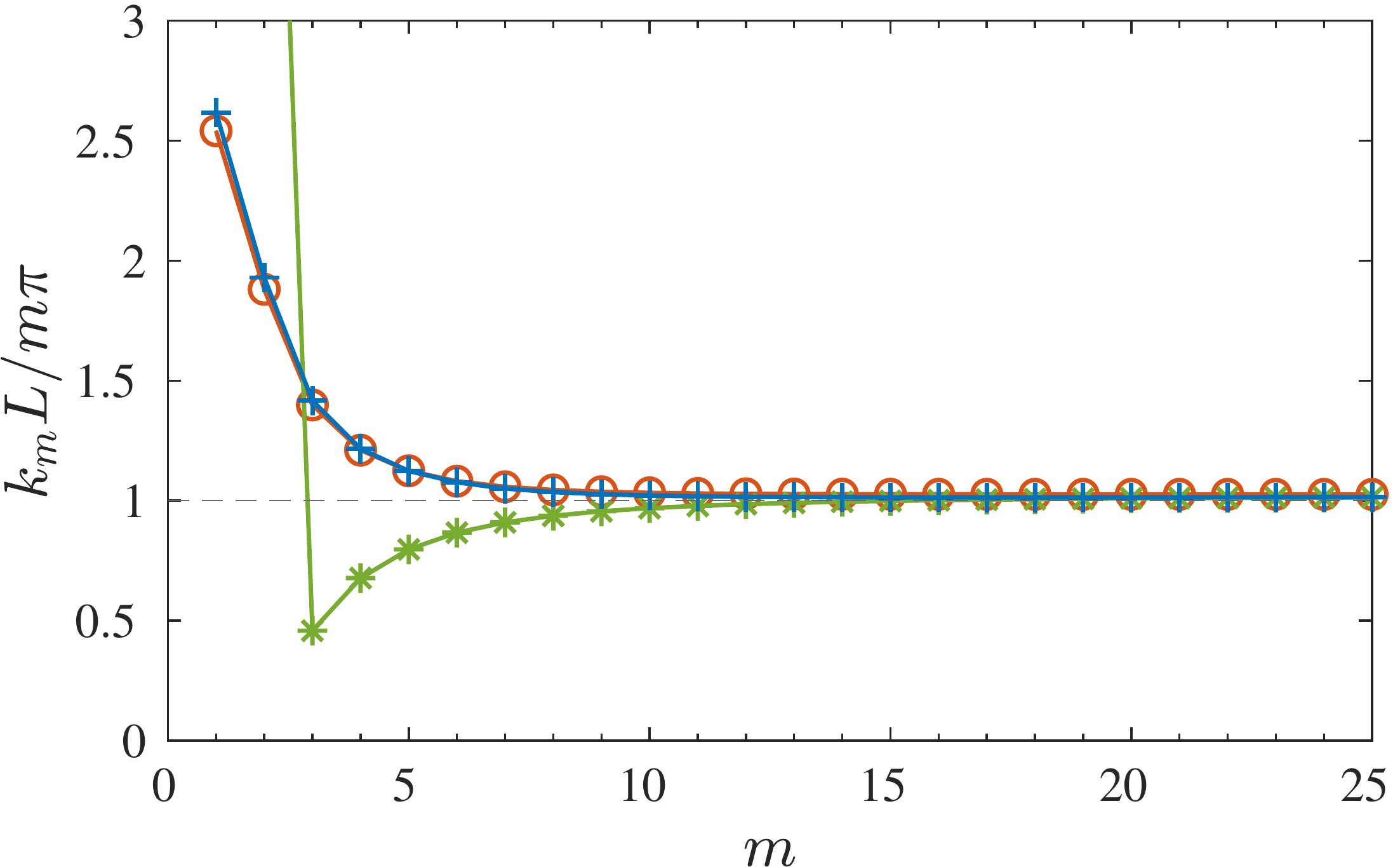}
	\caption{The ratio between the numerically extracted momentum $k_m$ of the spin-like mode $m$ and the corresponding analytical expectation for the uniform condensate $m\pi/L$. The red circles denote modes at $V_b/\mu=0$, the green stars at $V_b/\mu=1$ and blue crosses show $V_b/\mu=3$. The momentum for the density-like branch is indistinguishable from the spin-like one at $V_b/\mu=3$. Connection lines are a guide for the eye.}
	\label{fig:Split}
\end{figure}

In order to build dispersion relations of the Bogoliubov modes we need to associate a momentum value with each mode. 
This poses a non-trivial question for the trapped system as there is no translational symmetry in this case and the modes can never be eigenstates of the momentum operator. 
Following Refs. \cite{PhysRevA.89.053601,PhysRevA.88.043606} we introduce the longitudinal momentum of each mode as an expectation value of the squared momentum operator
\begin{equation}\label{eq:bog-mom}
k_m \equiv \sqrt{\langle m | k_x^2 | m \rangle} = \sqrt{\frac{\int d\mathbf{k} k_x^2 \left[|\tilde u_m (\mathbf{k})|^2 + |\tilde v_m (\mathbf{k})|^2 \right]}{\int d\mathbf{k} \left[|\tilde u_m (\mathbf{k})|^2 + |\tilde v_m (\mathbf{k})|^2 \right]}},
\end{equation}
where $\tilde{u}_m(\mathbf{k})$ and $\tilde{v}_m(\mathbf{k})$ are Fourier transforms of the $m$'th Bogoliubov mode $u_m\rr$ and $v_m\rr$. 
The momentum values obtained from Eq.~(\ref{eq:bog-mom}) can be compared to the idealized case of a uniform system of length $L$, where the eigenmodes of the momentum operator are distributed as $k_m = m \pi/L$ with $m\in\mathbb{Z}$. 
The results of such a comparison are presented in Fig.~\ref{fig:Split}. They show reasonable agreement especially for higher modes.
It is worth noticing that the momentum distribution of Bogoliubov modes is practically the same in low- and high-barrier regions. 
It is noticeably different only for spin-like modes and only for intermediate barrier heights $V_b/\mu \sim 1$.
In the discussions below we present the extracted momentum of the Bogoliubov modes in units of $\pi/L$ for easier interpretation.

\begin{figure}[tbp]
    \centering
    \includegraphics[width=\linewidth]{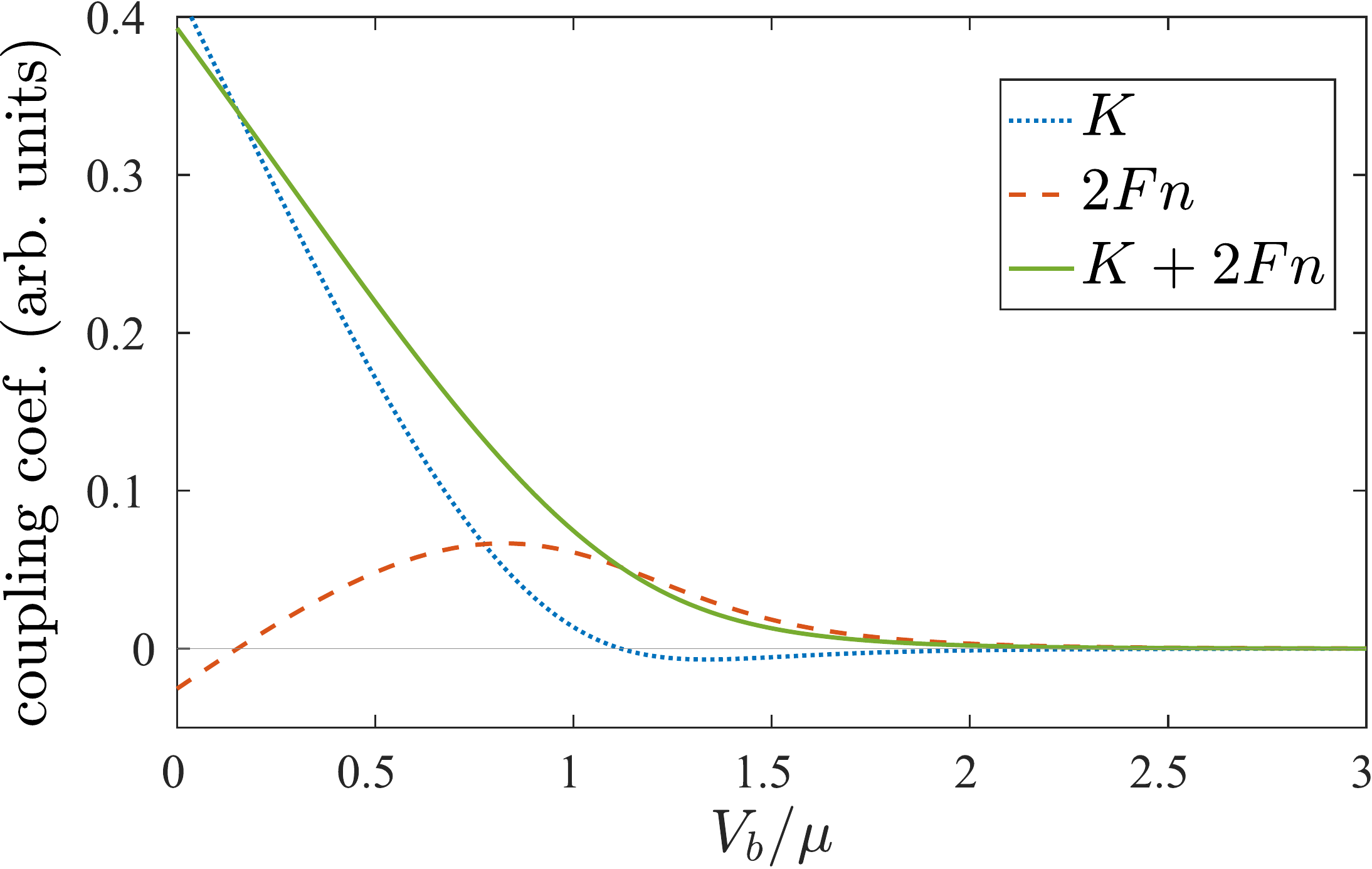}
    \caption{Linear ($K$) and nonlinear ($F n$) coupling terms, and their combination $K+2Fn$, which determine the stability of the ground state. 
    The quantities are shown as functions of the barrier height in the trapped system.
    }
    \label{fig:KF}
\end{figure}

To enable a comparison with the analytical predictions of the previous section it is necessary also to reliably approximate the one-dimensional model coefficients $g_{1D}$, $K$, and $F$, as well as the one-dimensional particle density $n$. 
To this end we calculate the antisymmetric first excited state $\psi_e$ of the Eq.~(\ref{eqn:GPE}), which is done by imaginary time evolution of an initial function with odd symmetry across the barrier.
Using two stationary states $\psi_g$ and $\psi_e$ all the necessary coefficients can be approximated directly (see \hyperref[app:coeffs]{Appendix} for details), which allows us to compare the analytical dispersion relations (\ref{eq:branch1}) and (\ref{eq:branch2}) to the numerical results without any fitted parameters. 
But before proceeding with such comparisons let us first analyze the obtained coefficients $K$ and $F$ to see if the one-dimensional model is valid for the trapped system under consideration.
The values of these coefficients depending on the ratio $V_b/\mu$ are shown in Fig.~\ref{fig:KF}. 
One can clearly see that there is a region with $K<0$. 
In this region the one-dimensional model would produce unphysical results if only linear coupling is considered. 
At the same time the characteristic combination $K + 2 F n$ appears to be always positive and monotonically decreasing. 
We can therefore conclude that the model based on Eq.~(\ref{eqn:1DC-GPE}) is applicable for our system in a wide range of barrier heights.

\begin{figure}[tbp]
	\includegraphics[width=\linewidth]{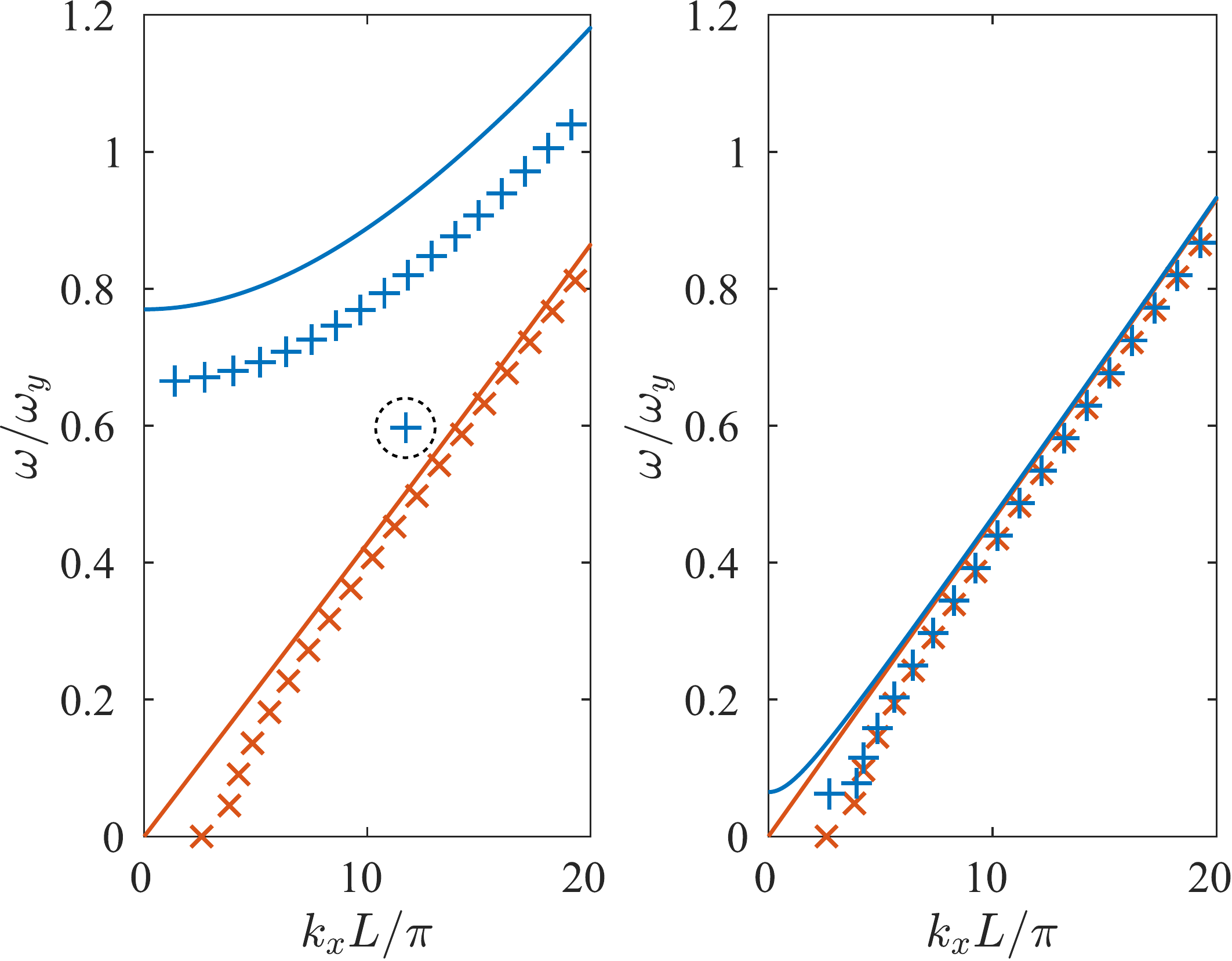}
	\caption{Dispersion relations for  $V_b/\mu =1$ (left panel) and $V_b/\mu = 2.5$ (right panel). Symbols denote the numerically obtained Bogoliubov modes of the trapped system. Solid lines show the corresponding analytical predictions of Eqs.~(\ref{eq:branch1}) and (\ref{eq:branch2}). 
	The red line and the symbols ``$\times$'' show the gapless density-like excitations, the blue line and the symbols ``$+$'' correspond to the gapped spin-like excitations. The dotted circle marks the location of the anomalous lowest spin-like mode discussed in the text.
	}
	\label{fig:Dispersion_examples}
\end{figure}

We now have all the necessary ingredients to analyze the dispersion relations of the coupled trapped condensates and compare them to the analytical formulas.
In Fig.~\ref{fig:Dispersion_examples} we show the results of such a comparison for two different values of the barrier height. 
We see that the general behavior of dispersion curves is adequately reproduced by the analytical model.
The main discrepancy, which is more pronounced in low- and intermediate-barrier regions, is the gap size for the spin-like modes.
The reason for this discrepancy is the same as that of a usual two-mode model. 
It originates from the fact that in a low-barrier region the spatial distribution of the lowest spin-like excitation is quite different from the shape of the antisymmetric excited state $\psi_e$, which was used to build the analytical model. 
More details on this effect can be found in Ref.~\cite{PhysRevA.95.023627}.
Other features of the dispersion curves, such as the slope of the density-like branch (the speed of sound) and the curvature of the spin-like branch, are reproduced much more accurately.

\section{Self-localization of spin-like modes}\label{sec:Modes}

One may notice in Fig.~\ref{fig:Dispersion_examples} (left panel) that the lowest spin-like mode deviates significantly from the corresponding branch of the dispersion relation. 
This mode, which is in fact two degenerate modes, shows considerably lower frequency and higher momentum values than expected from the smooth dispersion curve.
The existence of such anomalous modes is specific for the intermediate-barrier region $V_b/\mu \sim 1$.

\begin{figure}[tbp]
	\includegraphics[width=\linewidth]{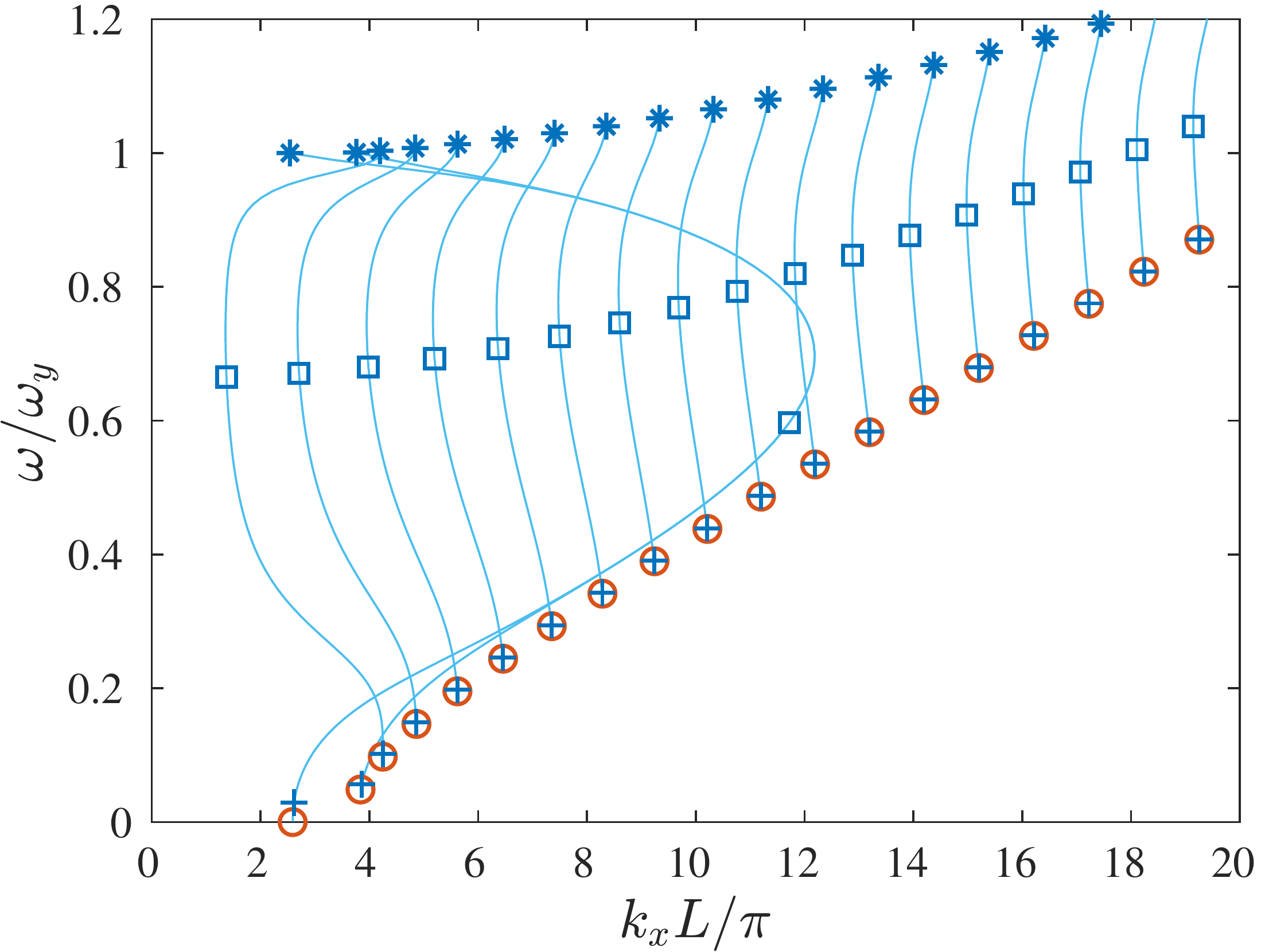}
	\caption{Frequency and momentum of the spin-like Bogoliubov modes depending on the barrier height. Blue symbols show the snapshots of this dependence for three specific barrier heights: $V_b/\mu=0$ (stars), $V_b/\mu=1$ (squares), $V_b/\mu=3$ (crosses). Red circles show the dispersion of symmetric Bogoliubov modes at $V_b/\mu=3$ for comparison. 
	Solid lines show the trajectory of each mode with changing barrier height $V_b$.
	}
	\label{fig:Disp}
\end{figure}

To further analyze the behavior of the lowest spin-like Bogoliubov excitations we trace the frequency and momentum of several modes of this type depending on the barrier height. 
The resulting trajectories are shown in Fig.~\ref{fig:Disp}.
As previously mentioned, a very simple model based on a finite uniform system suggests that momentum distribution of the modes is independent of the barrier height. 
From Fig.~\ref{fig:Disp} we see that higher excitations follow that prediction in general. 
The behavior of the lowest spin-like modes is however considerably different. 
Depending on the barrier height the two lowest modes may acquire a large momentum and become degenerate in both energy and momentum. 
Several other modes have their momentum considerably reduced in the same region.
The last observation is even more apparent from Fig.~\ref{fig:Split}, where we see that the entire spectrum of momentum values (except the lowest two modes) is considerably shifted downwards in the case of $V_b/\mu=1$.

\begin{figure}[tbp]
    \centering
    \includegraphics[width=\linewidth]{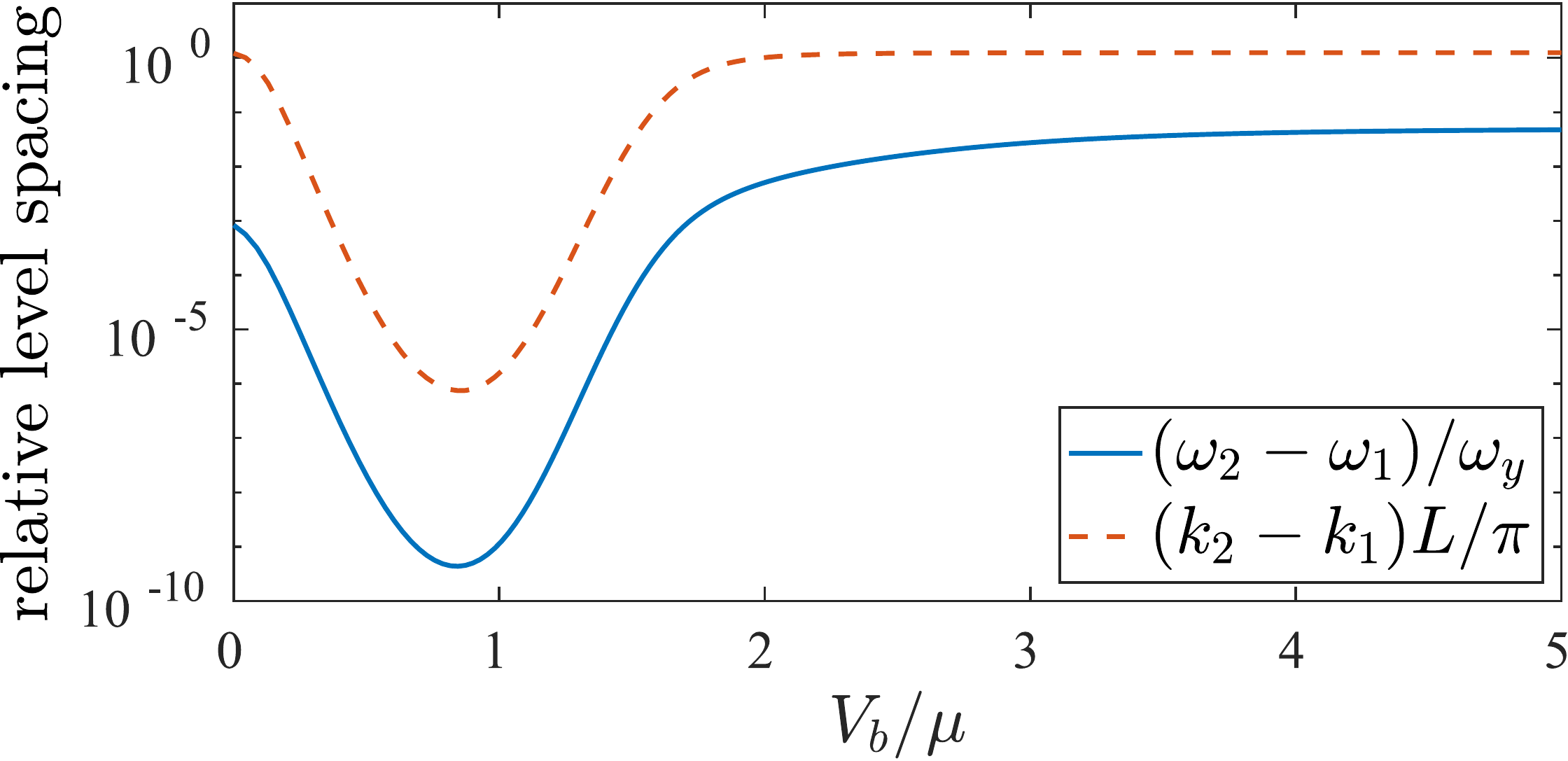}
    \caption{Relative frequency and momentum spacing between the two lowest spin-like modes as a function of barrier height. Note the logarithmic scale of the vertical axis. 
    }
    \label{fig:Degeneracy}
\end{figure}

In order to quantify the observed degeneracy we calculate the frequency and momentum difference between the two lowest spin-like modes (see Fig.~\ref{fig:Degeneracy}). 
Both these quantities show a pronounced minimum when the barrier height is of the same order as the chemical potential. 
The minimal values for both frequency and momentum spacings are reached simultaneously at $V_b/\mu \approx 0.87$. 
This value also corresponds to a maximal momentum of the two modes.

\begin{figure}[tbp]
    \centering
    \includegraphics[width=\linewidth]{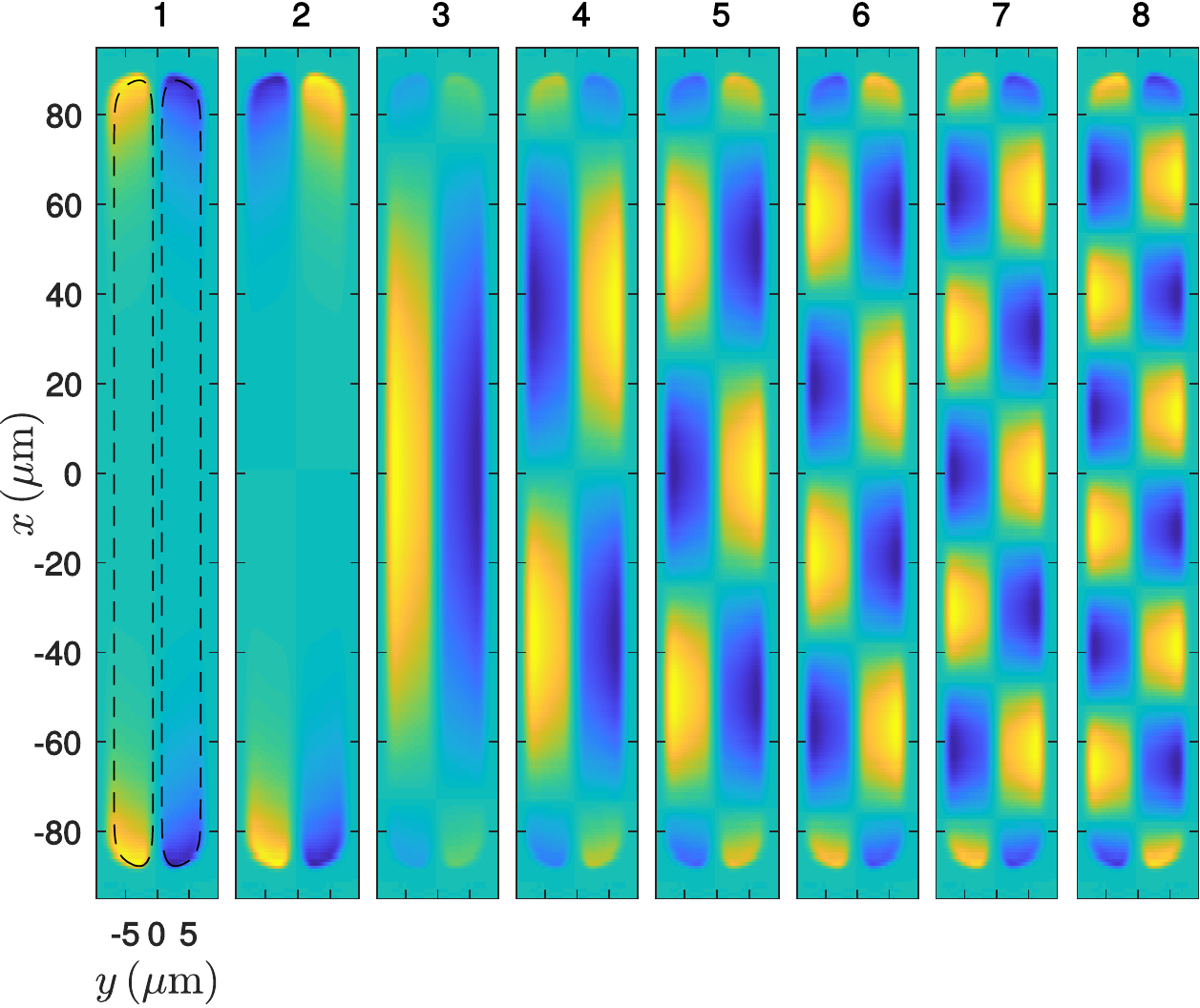}
    \caption{Lowest spin-like modes of the Bogoliubov spectrum for $V_b/\mu = 1$. The displacement from the equilibrium particle density $\delta n = \psi_g(u+v^*)$ is shown. Color represents the amplitude and sign of such a displacement. The dashed line on the first panel shows the $1/e^2$ isosurface of the ground-state density $|\psi_g|^2$.}
    \label{fig:modes}
\end{figure}

To get a better insight into anomalous behavior of the lowest spin-like modes we show in Fig.~\ref{fig:modes} spatial distributions of several lowest modes for $V_b/\mu = 1$. 
The two lowest ones, which are degenerate, appear to be also tightly localized at the edges of the condensate.
Such states cannot be characterized by reasonably well defined momentum values and Eq.~(\ref{eq:bog-mom}) rather represents the uncertainty of the momentum for them, which explains the high values observed in Fig.~\ref{fig:Disp}.
Other spin-like modes shown in Fig.~\ref{fig:modes} contain a node in their spatial distribution located close to the edges. 
This ensures orthogonality of the modes, but also may be responsible for the reduction of the corresponding momentum values, seen in Fig.~\ref{fig:Disp}.

\begin{figure}[tbp]
    \centering
    \includegraphics[width=\linewidth]{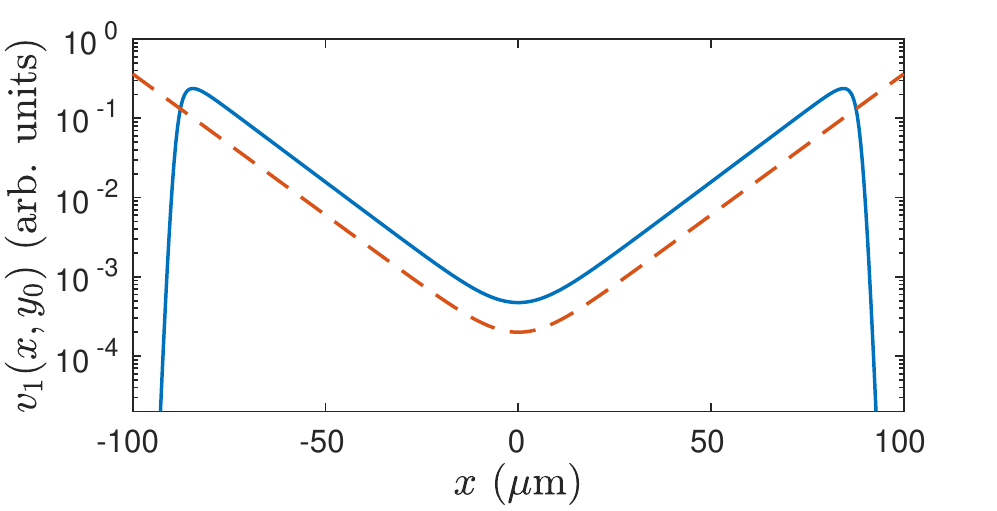}
    \caption{The solid blue line shows the lowest spin-like mode as a function of the $x$-coordinate at a fixed $y_0$ close to the minimum of one of the potential wells. The red dashed line shows the result of Eqs.~(\ref{eq:exp_mode}) and (\ref{eq:kappa}). Note the logarithmic scale of the vertical axis. The presented mode corresponds to $V_b/\mu = 1$.}
    \label{fig:mode_shape}
\end{figure}

Although the observed anomalous modes do not possess a well-defined momentum, they can still be related to the dispersion relation of spin-like modes in the homogeneous system (\ref{eq:branch2}).
Formally, this dispersion relation may have a real-frequency solution also with an imaginary value of the wavenumber $k=\ii\varkappa$. 
This implies exponentially growing solutions, which is certainly unphysical in an infinite system, but may be possible if the system is finite. 
From the spatial distribution of the first anomalous Bogoliubov mode $v_1$ shown in Fig.~\ref{fig:mode_shape} we see that away from the edges its $x$-dependence has a form of real exponentials
\begin{equation}
    v_1 \propto \ee^{\varkappa x} + \ee^{-\varkappa x},
    \label{eq:exp_mode}
\end{equation}
which indeed corresponds to the imaginary-wavenumber solution of the homogeneous system. 
Similarly, the second anomalous mode behaves as the difference of two exponentials with the same factor $\varkappa$.
Furthermore, it is possible to perform a more quantitative comparison with the analytical dispersion relation (\ref{eq:branch2}). 
To this end, we derive the exponential factor $\varkappa$ from Eq.~(\ref{eq:branch2}) by comparing the zero-wavenumber solution $\omega_2(0)$ and the imaginary-wavenumber solution $\omega_2(\ii\varkappa)$.
We get
\begin{equation}
    \varkappa^2 \approx M \frac{\omega_2^2(0)-\omega_2^2(\ii\varkappa)}{ g_\mathrm{1D} n + 2 K + 4 F n}.
    \label{eq:kappa}
\end{equation}
To perform a comparison with the numerically obtained spectrum we take the lowest normal mode frequency and the anomalous mode frequency as $\omega_2(0)$ and $\omega_2(\ii\varkappa)$ respectively.
From the results presented in Fig.~\ref{fig:mode_shape} we find a surprisingly good agreement of the above estimate with the behavior of the actual numerical solutions for the finite system.

To the best of our knowledge, the observed self-localization and mode degeneracy have not been analyzed previously in coupled atomic BECs. 
However, several similar phenomena were mentioned in previous theoretical studies.
In Ref.~\cite{PhysRevA.100.033601} the degenerate Bogoliubov modes are shown to be responsible for the multi-frequency plasma oscillations in long bosonic Josephson junctions. 
Edge-localized modes also show an apparent similarity with bending modes in the quasiparticle spectrum associated with a vortex line in a trapped BEC \cite{PhysRevLett.101.020402}.
Another similar self-localization behavior was also observed for tunneling currents in superconducting Josephson junctions \cite{PhysRev.164.538} and attributed to the Meissner effect. 
In all these examples the self-localized degenerate modes seemingly appear as an edge effect resulting from the elongated geometry of the system.
Further analysis of edge-localized solutions, especially concerning transitions between real and imaginary wavenumbers, and an extension to other trap potentials, remain interesting open questions for future studies.

\section{Conclusions}
In the present work we have performed an analytical and numerical study of the linear collective excitation spectrum in coupled elongated Bose-Einstein condensates.
The developed analytical model describes the dispersion of Bogoliubov quasiparticles in homogeneous condensates. 
The proposed approach is an extension of the well-known two-mode approximation. 
It offers the advantage that a direct comparison to numerical Gross-Pitaevskii simulations is possible without fitting parameters.
We have performed such a simulation for a realistic trapped system and compared the frequency and momentum spectra of quasiparticle excitations with analytical predictions. 
Such a comparison shows a reasonable agreement, taking into account the limitations of our analytical approach, mostly inherited from the two-mode approximation.

Our results also reveal anomalous behavior of the lowest spin-like excitations in the region of intermediate barrier heights $V_b/\mu \sim 1$.
The two lowest spin-like modes become degenerate and tightly localized at the edges of the condensate.
Such self-localization also leads to high momentum values obtained for these modes, which is mainly attributed to the imprecise definition of the mode momentum.
The self-localization of the lowest modes appears due to the finite size of the system and can be covered by our analytical model by inclusion of imaginary-wavenumber solutions.
Higher excitations retain a spatial structure of plane waves and follow the predicted dispersion relation, showing no signs of coupling to the tunneling motion.
This provides an intuitive explanation of the results obtained in \cite{Bidasyuk_2018}, where it was shown that thermal fluctuations have almost negligible influence on the tunneling dynamics. 
Similar degenerate modes were also previously observed in theoretical studies of harmonically-trapped condensates \cite{PhysRevA.100.033601}.
In that case however a large number of pairwise degenerate modes was observed. 
This indicates, in particular, that the number of anomalous modes may depend on the details of the longitudinal trap potential and one has to be cautious applying the results of the present work to other trap setups. 

The observed localized quasiparticle modes are a clear manifestation of a coupling between internal and mutual degrees of freedom in the two condensates. 
In realistic systems this coupling may lead to discrepancies of the tunneling dynamics from the Josephson model, which was observed in recent experiments  \cite{PhysRevLett.106.025302,PhysRevLett.120.025302} and also noticed in several numerical studies \cite{Bouchoule2005,PhysRevA.100.033601,Bidasyuk_2018}. 

\section*{Acknowledgements}

The authors are thankful to Michael Weyrauch for useful discussions and comments on the manuscript.

\appendix*

\section{Calculation of the coefficients in the 1D model}\label{app:coeffs}

Here we show how the coefficients of the one-dimensional model defined in Eqs.~(\ref{eq:g1d-coef}--\ref{eq:M-coef}) can be calculated using the solutions of the full system.
The main problem is that the separation of dimensions in Eq.~(\ref{eqn:1D-TMM-ansatz}) is only approximate in a real system, which means the functions $\chi_{1,2}$ are unknown. 
One could build an average guess of these functions, but a more useful approach is to rewrite the integrals of the coefficients, Eqs.~(\ref{eq:g1d-coef}--\ref{eq:M-coef}), in terms of known solutions of the full GPE (\ref{eqn:GPE}).
This is done by propagating Eq.~(\ref{eqn:GPE}) in imaginary time with an initial state prepared with the desired symmetry. 
Then we construct the following two functions
\begin{equation}
    \Phi_1(x,y,z) = \frac{\psi_g+\psi_e}{2}, \qquad \Phi_2(x,y,z) = \frac{\psi_g-\psi_e}{2}.
\end{equation}
These functions are by construction orthogonal and normalized to $N/2$. 
Each of them is also localized (mainly) in one of the potential wells.
We then assume that spatial dimensions can be separated as follows
\begin{equation}\label{eq:vsep}
\Phi_1 = \Psi_1(x)\chi_1(y,z), \quad \Phi_2 = \Psi_2(x)\chi_2(y,z).
\end{equation}
The (unknown) functions $\Psi_{1,2}$ and $\chi_{1,2}$ are assumed to have the following properties:
\[
\iint\dd{y} \dd{z} \chi_1\chi_2 = 0, \quad \iint\dd{y}\dd{z} \chi_1^2 = \iint\dd{y}\dd{z} \chi_2^2 = 1,
\]

\[
\Psi_1(x) = \Psi_2(x), \quad \int\dd{x} \Psi_1^2 = \frac{N}{2}.
\]
Such a decomposition satisfies the ansatz (\ref{eqn:1D-TMM-ansatz}) used in our one-dimensional model.
The above assumptions are sufficient to rewrite the integrals (\ref{eq:g1d-coef}--\ref{eq:M-coef}) in terms of the known functions $\Phi_1$ and $\Phi_2$. 
We get
\begin{align*}
	g_\mathrm{1D} &= \frac{4g}{N^2} \iint\dd{y} \dd{z} \left(\int \dd{x} \Phi_1^2\right)^2,\\
	K &= - \frac{2}{N} \iiint\dd{x}\dd{y}\dd{z} \left(-\frac{\hbar^2}{2m} \Phi_1 \laplacian_{y,z} \Phi_2 + V_{\perp} \Phi_1 \Phi_2 \right), \\
	F &= - \frac{4g}{N^2} \iint\dd{y}\dd{z} \left(\int \dd{x}\Phi_1^2\right) \left(\int \dd{x}\Phi_1\Phi_2\right), \\
	I &= \frac{4g}{N^2} \iint\dd{y}\dd{z} \left(\int \dd{x}\Phi_1\Phi_2\right)^2. 
\end{align*}
The one-dimensional density $n$, which enters the dispersion relations (\ref{eq:branch1}) and (\ref{eq:branch2}) can be estimated as follows:
\[
n = \max\limits_x \iint\dd{y} \dd{z} \Phi_1^2.
\]
Naturally, such an estimate is only valid if the condensate is mostly uniform in the $x$ dimension, which is the case for the trap potential (\ref{eqn:pot}) considered in the present work.

We stress that all the above expressions are straightforwardly calculated from the numerically obtained stationary solutions $\psi_g$ and $\psi_e$. 
They also provide a reasonable approximation if the separation of dimensions (\ref{eq:vsep}) is only approximate, which is always the case for our finite system.

\bibliography{refs}

\end{document}